\documentclass{iaus}
\usepackage{graphicx}

\title[Properties of Extragalactic GMCs] 
{The Resolved Properties of Extragalactic Giant Molecular Clouds}

\author[Alberto D. Bolatto et al.]   
{Alberto D. Bolatto$^1$, Adam K. Leroy$^2$,
  Erik Rosolowsky$^3$, Fabian Walter$^2$, \and Leo
  Blitz$^4$}

\affiliation{$^1$Department of Astronomy, University of Maryland, College Park, MD 20742,
 USA\\ email: {\tt bolatto@astro.umd.edu} \\[\affilskip]
$^2$Max-Planck-Institut f\"ur Astronomie, D-69117
Heidelberg, Germany \\[\affilskip]
$^3$ Department of Mathematics, Statistics, and Physics,
University of British Columbia at Okanagan, Kelowna, B.C. V1V 1V7,
Canada \\[\affilskip]
$^4$Department of Astronomy, University of California, Berkeley, CA94720, US
A}

\pubyear{2008}
\volume{255}  
\pagerange{119--126}
\jname{Low-Metallicity Star Formation: From the First Stars to Dwarf Galaxies}
\editors{L.K. Hunt, S. Madden \& R. Schneider, eds.}
\begin{document}

\maketitle

\begin{abstract}
Giant molecular clouds (GMCs) are the major reservoirs of molecular
gas in galaxies, and the starting point for star formation. As such,
their properties play a key role in setting the initial conditions for
the formation of stars.  We present a comprehensive combined
inteferometric/single-dish study of the resolved GMC properties in a
number of extragalactic systems, including both normal and dwarf
galaxies. We find that the extragalactic GMC properties measured
across a wide range of environments, characterized by the Larson
relations, are to first order remarkably compatible with those in the
Milky Way.  Using these data to investigate trends due to galaxy
metallicity, we find that: 1) these measurements do not accord with
simple expectations from photoionization-regulated star formation
theory, 2) there is no trend in the virial CO-to-H2 conversion factor
on the spatial scales studied, and 3) there are measurable departures
from the Galactic Larson relations in the Small Magellanic Cloud ---
the object with the lowest metallicity in the sample --- where GMCs
have velocity dispersions that are too small for their sizes. I will
discuss the stability of these clouds in the light of our recent
far-infrared analysis of this galaxy, and I will contrast the results
of the virial and far-infrared studies on the issue of the CO-to-H2
conversion factor and what they tell us about the structure of
molecular clouds in primitive galaxies.
\end{abstract}

\section{Introduction}

There is an emerging body of evidence suggesting that the formation of
Giant Molecular Clouds (GMCs) provides the regulating step in
transforming gas into stars in galaxies (\cite[Leroy et
al. 2008]{LEROY08}; see also Elias Brinks' contribution in these
proceedings).  Moreover, GMC properties set the initial conditions for
protostellar collapse, and likely play a determining role at setting
the initial stellar mass function (\cite[McKee \& Ostriker
2007]{MCKEE07}). However, there is a dearth of data on the properties
of GMCs in other galaxies, particularly low metallicity galaxies. Here
we introduce a systematic study of GMCs across dwarf galaxies, and
compare their properties with those measured in large galaxies and in
the Milky Way. The full analysis is discussed by \cite[Bolatto et
al. (2008)]{BOLATTO08}.

Studies of GMCs in the Milky Way (\cite[Solomon et al. 1987; Heyer et
al. 2004]{SOLOMON87,HEYER04}) find that GMCs are in approximate virial
equilibrium, and obey uniform scaling relations commonly known as
Larson laws (\cite[Larson 1981]{LARSON81}). These originate in the
compressible supersonic magnetohydrodynamic turbulence (also known as
Burger's turbulence) observed in the interstellar medium
(\cite[Elmegreen \& Scalo 2004]{ELMEGREEN04}).

\section{Results and Discussion}

How do extragalactic GMCs compare with those in the Galaxy?  We have
conducted a systematic study of spatially resolved extragalactic GMCs
in galaxies in the Local Group and beyond, using a combination of
interferometric and (for the Magellanic Clouds) single-dish CO
observations. All observations have been analyzed in the same manner,
using the {\tt CPROPS} algorithm described by \cite[Rosolowsky \&
Leroy (2007)]{ROSOLOWSKY07}.

\begin{figure}[b]
\begin{center}
 \includegraphics[width=5in]{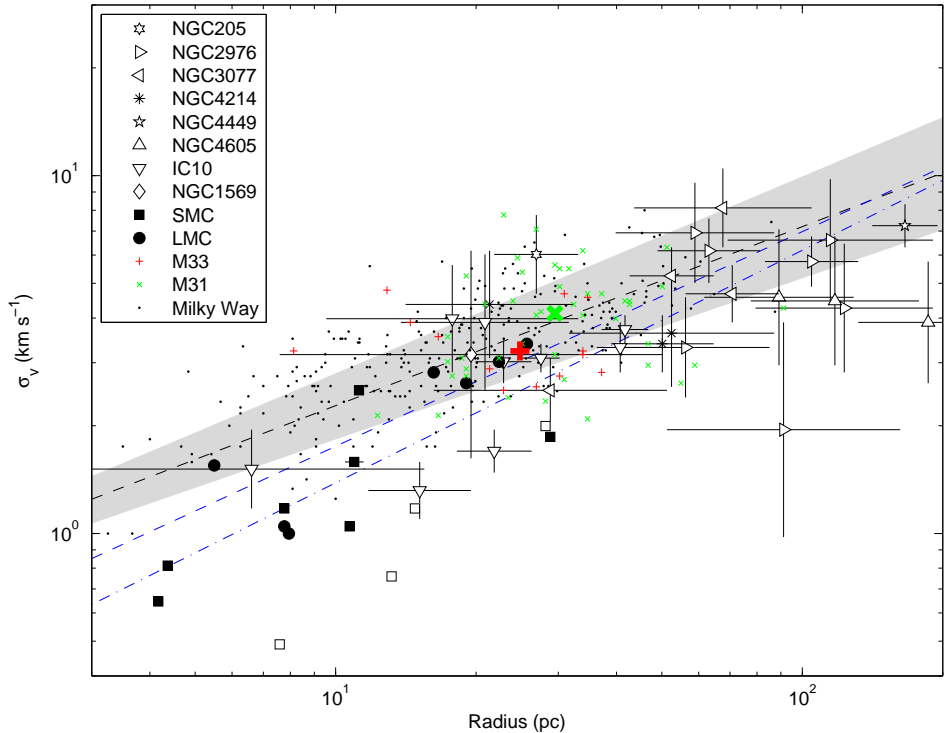} 

 \caption{Extragalactic size-line width relation. The gray region as
well as the small black points correspond to \cite[Solomon et
al. (1987)]{SOLOMON87}. The measurements for each galaxy are shown
with the corresponding error bars. For the Magellanic Clouds, white
and black symbols correspond to CO $1-0$ and CO $2-1$ measurements
respectively. The large symbols for M31 and M33 correspond to their
averages. The blue dashed and dot-dashed lines illustrate fits to all
galaxies and dwarf galaxies only, respectively \cite[(see Bolatto et
al. 2008 for details).]{BOLATTO08}}
 \label{fig1}
\end{center}
\end{figure}

Overall, we observe than the relations between size, velocity
dispersion, and luminosity observed fir extragalactic GMCs in our
sample are consistent with those determined in the Milky Way (Figure
\ref{fig1}). This result underscores that the Galactic Larson
relations provide a remarkably good description of CO-bright Giant
Molecular Clouds independent of their environment, at least in the
range of environments explored by this study (our lowest metallicity
galaxy, as well as the lowest metallicity galaxy in which CO emission
has been reliably detected, is the Small Magellanic Cloud).

Although the Larson relations are approximately Galactic there are
some significant departures. GMCs in dwarf galaxies tend to be
slightly larger than GMCs in the Milky Way, M~31, or M~33 for a given
CO luminosity or velocity dispersion. The largest departures occur in
the SMC. GMCs in dwarf galxies have average surface densities that are
a typically factor of two under that in the Galaxy ($\Sigma_{\rm
GMC}\approx170$ M$\odot$~pc$^{-2}$ for the Milky Way; \cite[Solomon et
al. 1987]{SOLOMON87}).  

In the case of the Small Magellanic Cloud and probably also for IC~10,
however, it seems that this explanation is not entirely viable since
the departures are much too large and central cloud extinctions would
be much too low. We suggest this is indirect evidence for large
molecular envelopes faint in CO. We will come back to this point in a
moment.

\begin{figure}[h]
\begin{center}
 \includegraphics[width=4.5in]{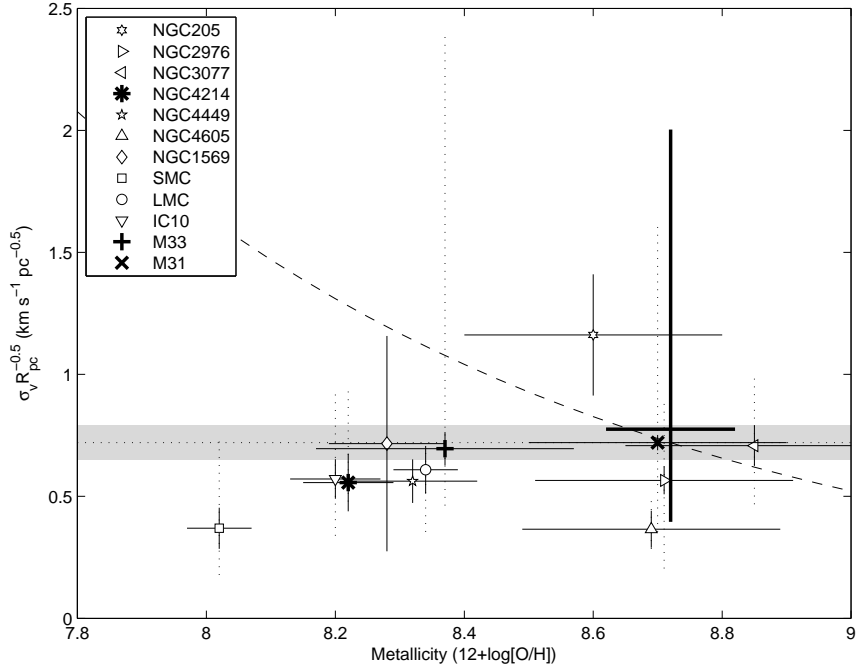} 
 \caption{Galaxy averages of surface density vs. metallicity for GMCs in our sample. Filled lines indicate mean uncertainty, dotted lines indicate full observed range. The thick lines correspond to the Milky Way (full range shown in the vertical). The dashed line shows the expectation from photoinization-regulated star formation theory \cite[(McKee 1989)]{MCKEE89}.}
   \label{fig2}
\end{center}
\end{figure}

\begin{figure}[t]
\begin{center}
 \includegraphics[width=4.5in]{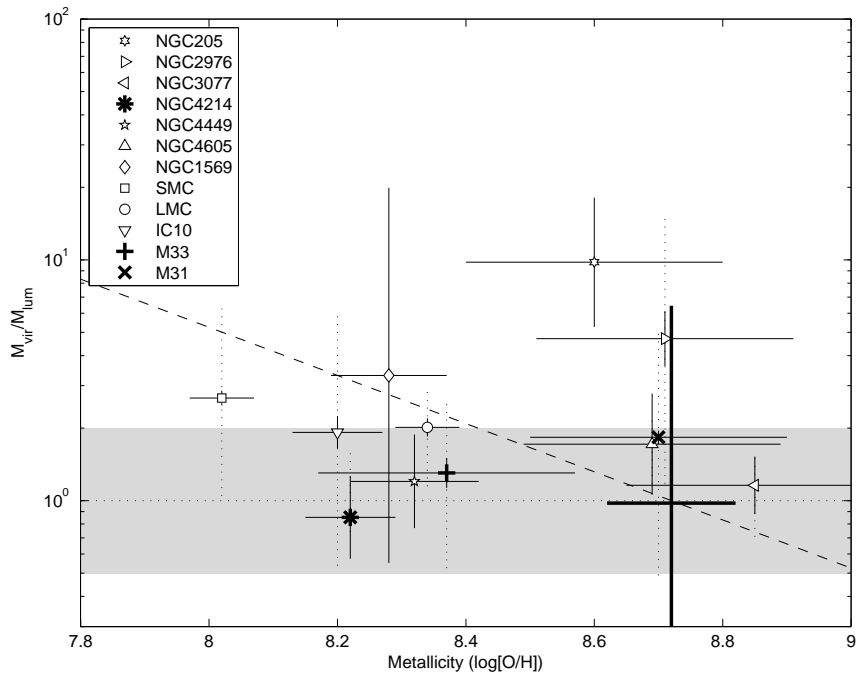} 
 \caption{CO-to-H$_2$ conversion factor, obtained as the ratio of virial over
luminous mass, versus metallicity for different galaxies in our sample. The gray line indicates the range in the Milky Way. The dashed line is the naive expectation $Z^{-1}$. Error bars are as described in Figure \ref{fig2}.}
   \label{fig3}
\end{center}
\end{figure}

Figure \ref{fig2} shows that our data does not show evidence for the
increase in GMC surface density with decreasing metallicity predicted
by photoionization-regulated star formation theory \cite[(McKee
1989)]{MCKEE89}. Figure \ref{fig3} shows that we do not see a
systematic clear increase in the ratio of the virial to luminous mass
(defined as the molecular mass obtained by using the Galactic
CO-to-H$_2$ conversion factor) as a function of metallicity in our
galaxies. In other words, we see an approximately constant CO-to-H$_2$
conversion factor in these GMCs. In fact, in the range of environments
probed (that is, normal and dwarf non starburst galaxies) the
properties of resolved CO-bright GMCs are very uniform. The departure
of the Small Magellanic Cloud from the Galactic ratio in Figure
\ref{fig3}, for example, is entirely ascribable to the fact that the
GMCs in this galaxy tend to be considerably smaller than GMCs in the
Milky Way and the relation between virial and luminous mass is weakly
dependent on mass \cite[(Solomon et al. 1987)]{SOLOMON87}.

These results stand in contrast to analyses that use the far-infrared
(either dust continuum or [CII] line emission) to trace molecular gas
in low metallicity environments \cite[(e.g., Madden et al. 1997; Leroy
et al. 2007)]{MADDEN97,LEROY07}. Studies of that type find a large
increase in the CO-to-H$_2$ conversion factor and large parcels of
CO-faint molecular gas. We suggest that the far-infrared and CO
observations can be simultaneously understood if we assume that bright
CO emission is only associated with the density peaks in metal-poor
environments. Observations that resolve these density peaks find, as
we show here, properties similar to Milky Way GMCs \cite[(see Heyer \&
Brunt 2004)]{HEYER04}. They miss, however, large envelopes of CO-faint
molecular gas. These envelopes would be CO-bright in objects of higher
metallicity.

Such envelopes may contain much of the molecular gas as low
metallicities \cite[(e.g., Leroy et al. 2007)]{LEROY07}.  Do they
participate in star-formation? In the case of the Small Magellanic
Cloud, for example, that appears necessary to preserve a normal star
formation efficiency. It is unclear, however, how translucent gas may
collapse and give rise to stars. Far-infrared observations of spectral
transitions as well as continuum with high spatial resolution will be
invaluable to understand the structure of the molecular gas in these
enviroments.

\end{document}